Multi-chain slip-spring simulations with various slip-spring densities

Yuichi Masubuchi[1,2*], Yuya Doi[2], and Takashi Uneyama[1,2]

[1]Center of Computational Science, and [2]Department of Materials Physics, Nagoya University, Nagoya 4648603, Japan.

*To whom correspondence should be addressed
mas@mp.pse.nagoya-u.ac.jp

**ABSTRACT**

Although it has been established that the multi-chain slip-spring (MCSS) model can reproduce entangled polymer dynamics, the effects of model parameters have not been fully elucidated yet. In this study, we systematically investigated the effects of slip-spring density. For the diffusion and the linear viscoelasticity, the simulation results exhibited universality. Namely, the results from the simulations with various slip-spring densities can be superposed with each other by the conversion factors for the bead number per chain, unit of length, unit of time, and modulus. The diffusion and the viscoelasticity were in good agreement with the literature data for the standard bead-spring simulations, including the molecular weight dependence. The universality among the MCSS simulations with various slip-spring density also held under mild shear if the slip-spring density was not significantly high. The results imply that the level of coarse-graining for the MCSS model can be arbitrarily chosen as in the Rouse model.

**INTRODUCTION**

Polymeric liquids exhibit entangled dynamics when the concentration and the molecular weight exceed the critical values[1]. The molecular simulations have revealed that the uncrossability between polymers plays an essential role[2,3]. However, the physics of entangled dynamics has not been fully elucidated yet. Nevertheless, lots of attempts have been made for the development of coarse-grained molecular models, in which the entangled polymer motion is reproduced by artificial restrictions to the polymer motion[4,5]. The representative one is the tube model[6], where the polymer motion is replaced by the single-chain dynamics confined in a tube-shaped constraint. Inspired by the remarkable success, several single-chain models have been proposed with the mean-field assumptions for the motion of surrounding chains[7–16]. Meanwhile, multi-chain models have also been developed to deal with the effects of neighboring chains explicitly[17–25].

The multi-chain slip-spring (MCSS) model is one of the multi-chain models for entangled polymers[20,22–24]. MCSS model deals with the motion of lots of Rouse chains connected by virtual springs. The virtual spring represents an entanglement between chains, and the anchoring point randomly slides along the chain, satisfying detailed balance. When the anchoring point comes to the chain end, the virtual spring is deleted by a certain probability. Vice versa, a new spring is introduced randomly to connect the chain end to the other Rouse bead.

For MCSS, owing to the nature of the Rouse chain, we expect that the number of Rouse beads between adjacent anchoring points of the virtual springs $N_{eSS}$ is a matter of arbitrariness. For unentangled polymers, the end-to-end distance and the longest relaxation time do not depend on the number of monomers replaced by the single Rouse segment, as long as the considered subchain is sufficiently large. MCSS would inherit this characteristic feature. As already discussed by Chappa et al.[22], MCSS exhibits the self-similarity for the chain statistics according to the construction of the free-energy, in which the inter-bead repulsive interaction eliminates the contribution of the virtual spring to the chain statistics. For the dynamics, as discussed for the tube model[6], the entangled polymer motion would be universal irrespectively of the definition of the Rouse segment if the intensity of dynamical constraint from the entanglement is consistent. Masubuchi[26] reported the dynamics of MCSS under steady shear with varying $N_{eSS}$, whereas the average number of anchoring points per chain, $Z_{SS}$, was fixed at 20. He demonstrated that the shear rate dependence of viscosity, relaxation time, diffusion, and chain dimension is not strongly dependent on $N_{eSS}$ in a range of $2.8 \leq N_{eSS} \leq 7.7$. However, no systematic study has been reported for the effect of $N_{eSS}$ on the dynamics of MCSS.

In this study, we investigated the chain dynamics for the melts of linear chains with various $N_{eSS}$ and $Z_{SS}$ to see the effects of $N_{eSS}$ mainly on the diffusion and viscoelasticity. Following the strategy for comparison among different multi-chain models[27], we first compared the molecular weight dependence of the diffusion constant $D$. We confirmed that the molecular weight dependence of $D$ could be superposed for the simulations with different $N_{eSS}$ values, and the result is in good agreement with the literature data by the standard bead-spring simulations. From this comparison, we determined the conversion factors for the bead number per chain $N$ and $D$. According to the conversion factor for $N$, we performed MCSS simulations under equilibrium for literature data of the bead-spring simulations to compare the mean-square displacement and the linear relaxation modulus. For the time development of these quantities, universality among the MCSS simulations with various $N_{eSS}$ also held, and semi-quantitative agreement with the bead-spring simulations was confirmed.

From the comparison, we determined the conversion factors for the unit of length, time, and modulus. Using the obtained conversion factors, we calculated the viscosity under shear and found that universality does not hold if $N_{eSS} < 2$. Details are shown below.

**MODEL AND SIMULATIONS**

Because the model and the simulation code are the same as those employed in the earlier studies[20,26–29], only a brief explanation is given below. In the MCSS simulations, a lot of Rouse chains are dispersed in the simulation box and connected via virtual springs, which are so-called slip-spring. The state valuables of the system are the position of the *k*-th Rouse beads on the *i*-th chain, $\{\mathbf{R}_{i,k}\}$, and the connectivity matrix of the slip-spring $\{S_\alpha\}$. The state of $\alpha$-th spring is defined by its anchoring points on the beads $(S_{\alpha,1}, S_{\alpha,2})$ and $(S_{\alpha,3}, S_{\alpha,4})$. This notation means that one of the anchoring points is located at the $S_{\alpha,2}$-th Rouse beads on the $S_{\alpha,1}$-th chain, and the other one is at the $S_{\alpha,4}$-th Rouse beads on the $S_{\alpha,3}$-th chain.

The total free energy of the system is written as

$$\frac{F}{k_B T} = \frac{3}{2b^2} \sum_{i,k} (\mathbf{R}_{i,k+1} - \mathbf{R}_{i,k})^2 + \frac{3}{2N_s b^2} \sum_\alpha (\mathbf{R}_{S_{\alpha,1},S_{\alpha 2}} - \mathbf{R}_{S_{\alpha,3},S_{\alpha,4}})^2$$
$$+ e^{\nu/kT} \sum_{i,k,j,l} \exp\left[-\frac{3}{2N_s b^2} (\mathbf{R}_{i,k} - \mathbf{R}_{j,l})^2\right] \quad (1)$$

Here, the first term in the right-hand side is the contribution from the chain connectivity, and $b$ is the average bond length between consecutive Rouse beads. The second term corresponds to the elastic energy of the slip-springs with the spring constant of $N_s b^2/3$ where $N_s$ is the parameter that determines the stiffness of the slip-springs. The third term is the soft-core repulsive interaction between all the pairs of the Rouse beads in the system, and $e^{\nu/kT}$ is the fugacity of the virtual springs. This term precisely eliminates the artificial contributions from the second term on the chain statistics, and it retains the Gaussian chain statistics. Following the previous studies, the parameter $N_s$ was fixed at 0.5 in the present study. Note that Vogiatzis et al.[30] mentioned that $N_s$ should be larger than unity. However, in our practice, $N_s = 0.5$ can preserve the unperturbed fractal chain dimension, as reported previously[20].

According to the free energy, $\{\mathbf{R}_{i,k}\}$ were numerically developed by a Langevin equation given by

$$\zeta \left(\frac{d\mathbf{R}_{i,k}}{dt} - \boldsymbol{\kappa} \cdot \mathbf{R}_{i,k}\right) = -\frac{\partial F}{\partial \mathbf{R}_{i,k}} + \boldsymbol{\xi}_{i,k}(t). \quad (2)$$

Here, $\zeta$ is the friction coefficient of a Rouse bead, $\boldsymbol{\kappa}$ is the velocity gradient tensor, and $\boldsymbol{\xi}_{i,k}(t)$ is the Gaussian random force that obeys

$$\langle \boldsymbol{\xi}_{i,k}(t)\rangle = \mathbf{0}, \qquad \langle \boldsymbol{\xi}_{i,k}(t)\boldsymbol{\xi}_{j,l}(t')\rangle = \frac{2kT}{\zeta}\delta_{ij}\delta_{kl}\delta(t-t')\mathbf{I} \tag{3}$$

Representing the sliding motion of slip-springs along the chain, $\{S_\alpha\}$ was changed according to the Monte Carlo scheme. For a numerical integration time step $\Delta t$, all the slip-springs attempt to hop along the chain. Such dynamics is described by a change of the connectivity matrix. For $\alpha$-th spring, the change $S_{\alpha,\lambda} \rightarrow S_{\alpha,\lambda\pm 1}$ ($\lambda = 2, 4$) occurred according to the cumulative probability written as

$$\psi_{\lambda\pm} = \frac{\Delta t}{\zeta_s}\left(1 - \tanh\frac{\Delta F_{\lambda\pm}}{2}\right) \qquad (\lambda = 2, 4) \tag{4}$$

Here, $\Delta F_{\lambda\pm}$ is the difference of the free energy according to the change of the connectivity, and $\zeta_s$ is the friction for the sliding.

In addition to the sliding, we dealt with the creation and annihilation of the slip-springs. At each chain end, a new virtual spring $\beta$ is created by the following probability.

$$\psi_+ = \exp\left[-\frac{3(\mathbf{R}_{S_{\beta,1},S_{\beta,2}} - \mathbf{R}_{S_{\beta,3},S_{\beta,4}})^2}{2N_s}\right] \tag{5}$$

Here, $(S_{\beta,1}, S_{\beta,2})$ represents the subjected chain end, and $(S_{\beta,3}, S_{\beta,4})$ is another bead, which is randomly chosen from the surrounding beads within a cut-off distance $r_c$. The total number of trials for the creation of slip-spring was

$$K = 4M\frac{4\pi}{3}r_c^3\rho\frac{\Delta t}{\zeta_s}e^{\frac{\nu}{kT}} \tag{6}$$

Here, $\rho$ is the density of the Rouse beads. Vice-versa, a slip-spring is annihilated when one of the anchoring points comes to the chain end with the following cumulative probability.

$$\psi_- = \frac{\Delta t}{\zeta_s} \tag{7}$$

The kinetics shown above fulfill the detailed balance conditions.

We numerically integrated the equation of motion for $\{\mathbf{R}_{i,k}\}$ according to the 2nd-order scheme[31]. The calculations were made for the dimensionless valuables, for which the unit of length, energy and time were chosen as the segment length $b$, the thermal energy $kT$, and the diffusion time of the single bead $\tau = \zeta b^2/kT$. The parameters for the slip-spring dynamics were chosen as $N_s = 0.5$, $\zeta = \zeta_s = 1$, $\Delta t = 0.1$, $r_c^2 = 1.29$, and $\rho = 4$. The chain number $M$ in the system was fixed at 200 for all the examined cases. Cubic simulation boxes were used with periodic boundary conditions. For shear deformations, Lees-Edwards boundary condition was employed. For statistics, 8 independent simulations from different initial configurations were made for the simulation runs at least 10 times longer than the longest relaxation time of the system.

**RESULTS AND DISCUSSION**

Figure 1 shows the entanglement molecular weight, $N_{\text{eSS}}$, defined as

$$N_{\text{eSS}} = \frac{NM}{2\langle M_{\text{SS}}\rangle} \quad (8)$$

Here, $M_{\text{SS}}$ is the total number of slip-springs in the system, and $\langle \cdots \rangle$ means the ensemble average. The factor 2 in the denominator means that there are two anchoring points for each slip-spring. In Figure 1, left panel, we plot $N_{\text{eSS}}$ as a function of $N$. By the model construction, $N_{\text{eSS}}$ decreases with increasing $e^{\nu/kT}$, reflecting an increase of $M_{\text{SS}}$. For each $e^{\nu/kT}$ value, $N_{\text{eSS}}$ slightly increases with increasing $N$, and approaches to a steady value for large $N$.

The $N$ dependence of $N_{\text{eSS}}$ is due to the contribution of intrachain slip-springs that explicitly depends on $N$. Here, we call a slip-spring as a intrachain slip-spring if both ends are attached to the same chain[32]. As discussed previously[20], the average number density of slip-springs, $\phi \equiv \langle M_{\text{SS}}\rangle/V$ (where $V$ is the volume) is given as

$$\phi = \rho e^{\nu/kT}\left[1 + \frac{2}{N}\sum_{k=1}^{N}\sum_{l=1}^{k-1}\left(\frac{N_s}{k-l+N_s}\right)^{3/2} + \left(\rho - \frac{N}{V}\right)\left(\frac{2\pi N_s b^2}{3}\right)^{3/2}\right] \quad (9)$$

If $N$ and $V$ are sufficiently large, we have

$$\phi \approx \rho e^{\nu/kT}\left[2\Omega(N_s) - 1 + \rho\left(\frac{2\pi N_s b^2}{3}\right)^{3/2}\right] \quad (10)$$

Here, $\Omega(N_s)$ is the following factor which represents the strength of the intrachain contribution.

$$\Omega(N_s) = \sum_{k=0}^{\infty}\left(\frac{N_s}{k+N_s}\right)^{3/2} = N_s^{3/2}\zeta(3/2, N_s) \quad (11)$$

Here, $\zeta(s,a)$ is the Hurwitz zeta function. Nevertheless, from eq 10, we have the following relationship.

$$\frac{1}{N_{\text{eSS}}^{\infty}} = \frac{2\phi}{\rho} = 2e^{\nu/kT}\left[2\Omega(N_s) - 1 + \rho\left(\frac{2\pi N_s b^2}{3}\right)^{3/2}\right] \quad (12)$$

Here, $N_{\text{eSS}}^{\infty}$ is the value of $N_{\text{eSS}}$ in the large $N$ limit. Eq 12 implies that $1/N_{\text{eSS}}^{\infty} \propto e^{\nu/kT}$, because $N_s$, $\rho$, and $b$ are constants. Substituting the employed parameters, we have $1/N_{\text{eSS}}^{\infty} \sim 8.8\, e^{\nu/kT}$. This relation is confirmed in Figure 1 right panel, where $N_{\text{eSS}}^{\infty}$, is shown as a function of $e^{\nu/kT}$ with the theoretical relation.

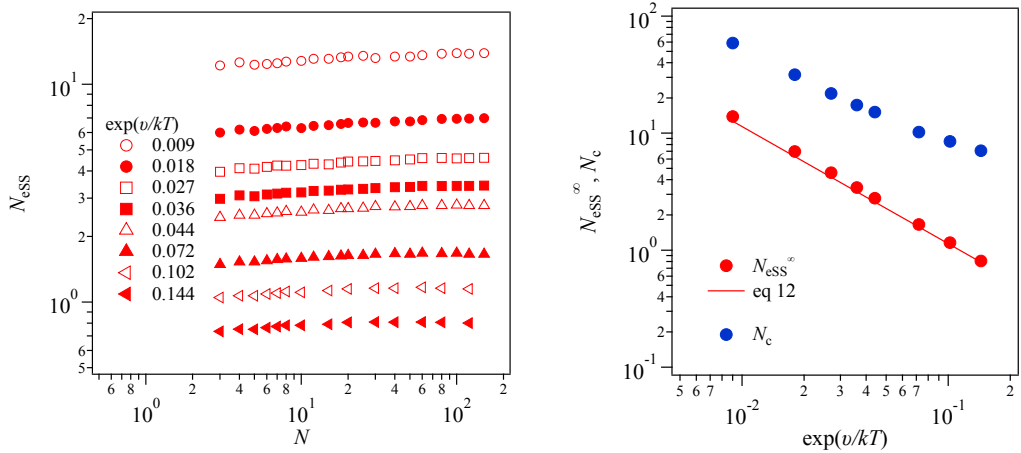

**Figure 1** Entanglement molecular weight $N_{eSS}$ as a function of the bead number per chain $N$ for various values of the slip-spring fugacity $e^{\nu/kT}$ (left), and the asymptotic value of $N_{eSS}$ for large-$N$ limit, $N_{eSS}^{\infty}$, and the critical molecular weight for the onset of entanglement, $N_c$, plotted against $e^{\nu/kT}$ (right). In the right panel, the line shows eq 12.

Figure 2 left panel shows the molecular weight dependence of the diffusion coefficient $D$. As previously reported[20,27], $D$ decreases with increasing $N$ showing both of unentangled and entangled behaviors. $D$ is suppressed as $e^{\nu/kT}$ increases because of the rise of $M_{SS}$. To demonstrate the transition between unentangled and entangled regime, Figure 2 right panel displays $DN^2$ for comparison to the tube theory, in which $D \propto N^{-2}$. In a small $N$ region, reflecting the Rouse behavior (i.e., $D \propto N^{-1}$), $DN^2$ is proportional to $N$. In contrast, in a large $N$ region, $DN^2$ decreases with increasing $N$, exhibiting that the reduction of $D$ is more significant than the prediction of the tube theory, as observed for experiments[32].

We define the critical molecular weight for the onset of entanglement, $N_c$, from the peak position of $DN^2$ shown in Figure 2 right panel. The $N_c$ values thus obtained are plotted against $e^{\nu/kT}$ in Figure 1, right panel (blue circle). Experimentally, the critical molecular weight of entanglement is in proportional to the entanglement molecular weight. We expected such a relationship between $N_c$ and $N_{eSS}^{\infty}$. This thought works reasonably well for small $e^{\nu/kT}$ values, where $N_{eSS}^{\infty}$ is sufficiently larger than unity. In this region, $N_c = 4.2 N_{eSS}^{\infty}$ as previously reported[27]. Meanwhile, when $e^{\nu/kT}$ is large and $N_{eSS}^{\infty}$ is small, the proportionality does not hold. $N_c$ deviates from the slope of -1 for large $e^{\nu/kT}$, where $N_{eSS}^{\infty}$ becomes smaller than 2. This behavior of $N_c$ implies that the dynamic constraint induced by the slip-springs becomes relatively weak when the slip-springs are densely connected to the chains. For instance, for the case of $e^{\nu/kT} = 0.144$, $N_c = 8.7 N_{eSS}^{\infty}$, demonstrating that the number of slip-springs per chain required for the onset of entanglement is more than double in comparison to the case with $e^{\nu/kT} = 0.009$ (where $N_c = 4.2 N_{eSS}^{\infty}$).

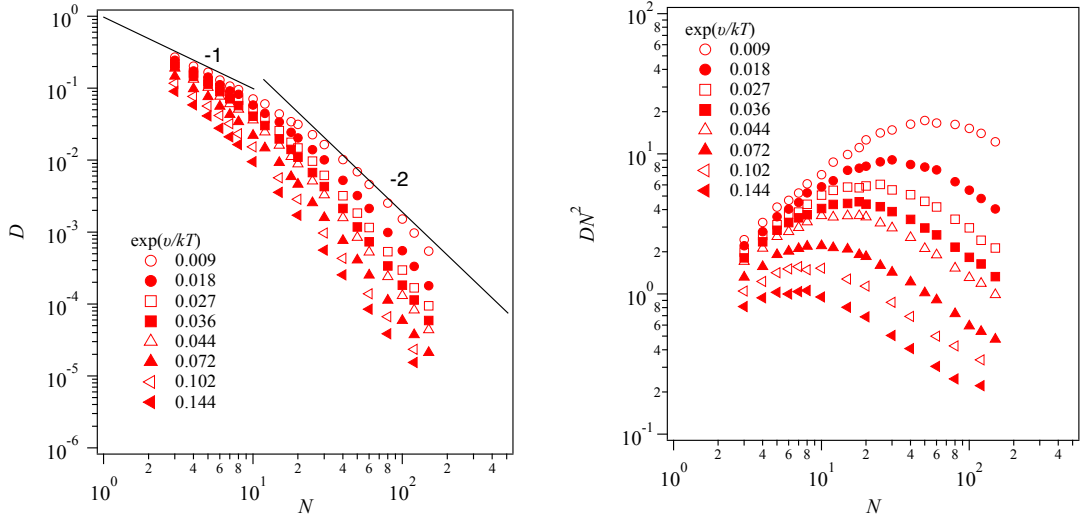

**Figure 2** Molecular weight dependence of the diffusion constant $D$ for various $e^{v/kT}$ values (left), and its normalized value concerning the reptation prediction $DN^2$ (right). The lines in the left panel indicate slopes of -1 and -2, which correspond to the Rouse and the reptation predictions, respectively.

The molecular weight dependence of $D$ is universal among the results with various $e^{v/kT}$, and the plots of $D(N)$ and $DN^2(N)$ can be superposed if the conversion factors for the bead number and the diffusion constant are adequately chosen. Figure 3 shows the conversion of $DN^2(N)$ to the literature data obtained for the standard bead-spring simulations [33–37]. The subscript KG stands for Kremer and Grest[33], who are the inventor of the model. Although the data for the bead-spring simulations are not available for the large $N$ region, we have confirmed a reasonable agreement for the unentangled and transitional behaviors.

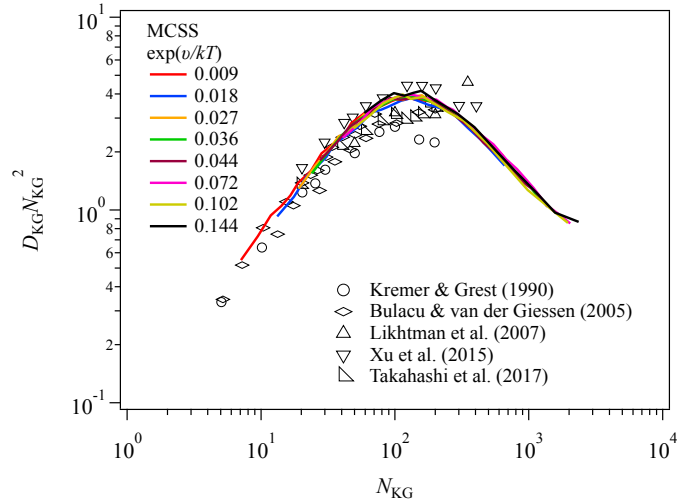

**Figure 3** The normalized diffusion constant to the tube prediction, $DN^2$, as a function of the bead number per chain $N$ for various $e^{v/kT}$ values converted to the results of the standard bead-spring simulations extracted from the literature [33–37]. The conversion factors for the MCSS simulations are summarized in Table I.

The plot shown in Figure 3 gives the conversion factors for $N$ and $D$, as summarized in Table I. Here, $N_{KG}$ and $D_{KG}$ are the bead number per chain and the diffusion constant for the bead-spring simulations. Because of the conversion relying on $N_c$, the relations between $e^{v/kT}$ and the conversion factors are not trivial. Namely, $N_{KG}/N$ is in proportional to $N_c$ rather than $N_{eSS}^{\infty}$.

We note that similar conversion can be attained for the molecular weight dependence of the longest relaxation time and the viscosity. However, we have a few practical disadvantages for the use of these viscoelastic quantities. These values cannot be obtained for each molecule due to the contribution of the inter-chain correlation[27]. Thus, the obtained values are with a relatively lower accuracy than the diffusion, for which we take ensemble average among the molecules. The other reason is the amount of accumulated results for the bead spring simulations. We can find several datasets in the literature as shown in Fig 3 for diffusion, but not that many for viscoelastic quantities. Nevertheless, as we shall discuss later, the conversion based on $N_c$ works well for the viscoelastic quantities.

**Table I** Conversion factors for MCSS models with various $e^{v/kT}$ values to the standard bead-spring model[33].

| $e^{v/kT}$ | $N_{KG}/N$ | $D_{KG}/D$ | $(\sigma_{KG}/b)^2$ | $\tau_{KG}/\tau$ | $G_{KG}/G$ |
|---|---|---|---|---|---|
| 0.009 | $2.3_6$ | $4.0_4 \times 10^{-2}$ | $3.0_2$ | $7.4_7 \times 10$ | $1.9_9 \times 10^{-1}$ |
| 0.018 | $4.3_9$ | $2.1_7 \times 10^{-2}$ | $5.6_3$ | $2.5_9 \times 10^2$ | $9.9_8 \times 10^{-2}$ |
| 0.027 | $6.3_5$ | $1.6_1 \times 10^{-2}$ | $8.1_3$ | $5.0_5 \times 10^2$ | $6.5_7 \times 10^{-2}$ |
| 0.036 | $8.0_0$ | $1.3_4 \times 10^{-2}$ | $10._2$ | $7.6_5 \times 10^2$ | $5.0_0 \times 10^{-2}$ |
| 0.044 | $9.1_8$ | $1.2_5 \times 10^{-2}$ | $11._7$ | $9.4_1 \times 10^2$ | $3.9_7 \times 10^{-2}$ |
| 0.072 | $13._6$ | $9.7_0 \times 10^{-3}$ | $17._4$ | $1.7_9 \times 10^3$ | $2.3_7 \times 10^{-2}$ |
| 0.102 | $16._3$ | $9.4_8 \times 10^{-3}$ | $20._9$ | $2.2_1 \times 10^3$ | $1.6_5 \times 10^{-2}$ |
| 0.144 | $19._6$ | $1.0_2 \times 10^{-2}$ | $25._1$ | $2.4_7 \times 10^3$ | $1.1_6 \times 10^{-2}$ |

Figure 4 displays snapshots of the MCSS chains that are equivalent to the standard bead-spring model with $N_{KG} = 512$ for various $e^{v/kT}$ values. The number of beads per chain decreases with an increase of $e^{v/kT}$, representing the increase of the molecular weight represented by the single MCSS bead as indicated by $N_{KG}/N$ in Table I.

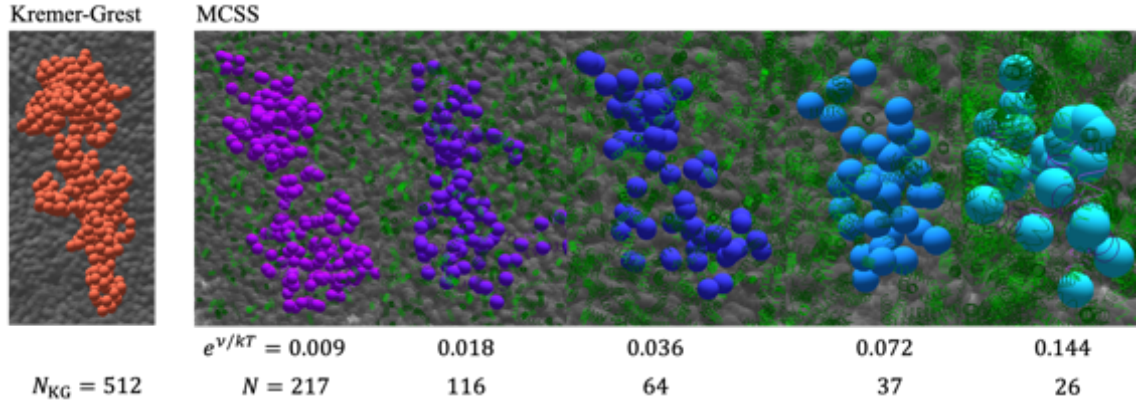

**Figure 4** MCSS chains representing the bead-spring chain with $N_{KG} = 512$ for various $e^{v/kT}$ values. The gray spheres are the surrounding molecules, and the green springs are the slip-springs.

To decompose the conversion factor for $D$ into that for the length and time, we compared the mean-square-displacement of the bead at the chain center, $g_1(t)$, with the literature data from the simulations of the standard bead-spring chains[33–38], as shown in Figure 5. As established earlier[6], $g_1(t)$ shows a few transitions at which the slope changes reflecting the following dynamics: the segment motion perpendicular to the tube $(g_1(t) \sim t^{1/2})$, the Rouse-like motion along the contour $(g_1(t) \sim t^{1/4})$, the overall diffusion along the contour $(g_1(t) \sim t^{1/2})$, and the normal diffusion $(g_1(t) \sim t^1)$. In the right panel, we plot $g_1(t) t^{-1/2}$ to clarify the transitions between different regions, following the earlier study[35]. For the examined cases, the deviation between the results is smaller than the scattering among the literature data for the bead-spring simulations.

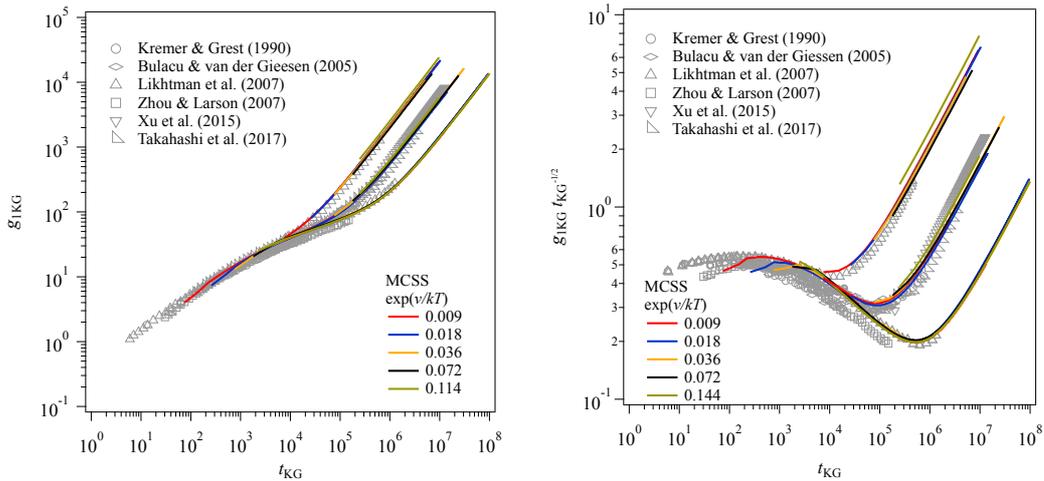

**Figure 5** Mean-square-displacement of the central bead, $g_1(t)$ (left) and its normalized value to the Rouse behavior, $g_1(t) t^{-1/2}$ (right), both converted to the results from the standard bead-spring simulations extracted from the literature[33–38] for $N_{KG}$ =100, 200, and 350, from left to right. The conversion factors are summarized in Table I.

The conversion factors thus obtained for the unit of length $(\sigma_{KG}/b)^2$ and time $\tau_{KG}/\tau$ are shown in Table I. Here, $\sigma_{KG}$ and $\tau_{KG}$ are the unit of length and time for the bead-spring model. Note that for all the examined cases, the relation $D_{KG}/D = (\sigma_{KG}/b)^2/(\tau_{KG}/\tau)$ is fulfilled. The conversion factor $(\sigma_{KG}/b)^2$ is in proportion to $N_{KG}/N$ as shown in Figure 6. This proportionality is expected from the Gaussian chain statistics. $\tau_{KG}/\tau$ is proportional to the square of $N_{KG}/N$ being consistent with the Rouse behavior of the subchain when $e^{\nu/kT}$ is small. In contrast, for large $e^{\nu/kT}$ values, $\tau_{KG}/\tau$ is proportional to $N_{KG}/N$. For this case, $N_{eSS}^\infty$ is smaller than 2 as seen in Figure 1, and the subchain does not exhibit the Rouse behavior due to the densely attached slip-springs.

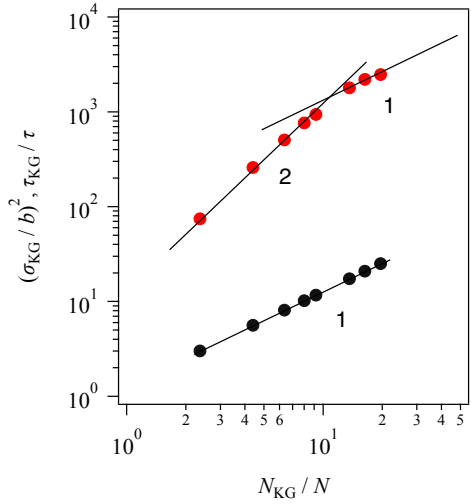

**Figure 6** Conversion factors for MCSS models to the standard bead-spring model for length $(\sigma_{KG}/b)^2$ (black circle) and time $\tau_{KG}/\tau$ (red circle) as functions of the conversion factor for the bead-number per chain $N_{KG}/N$. The lines indicate slopes of 1 and 2.

Figure 7 shows the comparison of the squared end-to-end distance $\mathbf{R}_{KG}^2$ between MCSS and the standard bead-spring model as a function of the bead number per chain $N_{KG}$. The MCSS results were converted by the conversion factors $(\sigma_{KG}/b)^2$ and $N_{KG}/N$ in Table 1. As reported previously, by the conversion factors obtained from the diffusion, the chain dimension is underestimated in comparison to the results from the bead-spring simulations[33,37,39,40]. The chain dimension in the MCSS models is Gaussian, and the downward deviation seen for the short chains is because $\mathbf{R}^2$ is proportional to the bond number $(N-1)$, not to the bead number $N$. In the long chain region, MCSS results with large $e^{\nu/kT}$ values exhibit a slight chain swelling, probably due to a flaw in our numerical implementation for the soft-core interaction. Nevertheless, the deviation is sufficiently smaller than the distribution shown by the error bars.

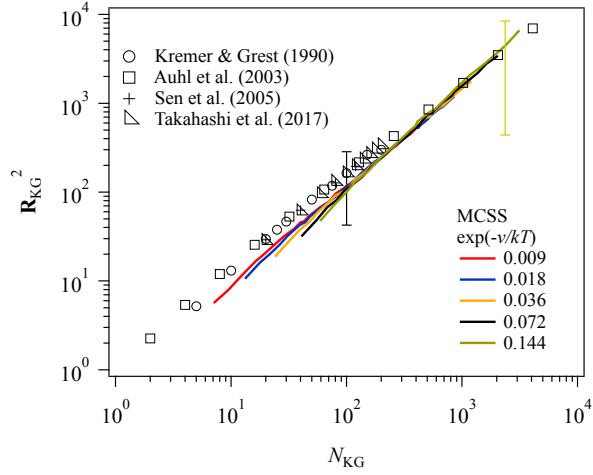

**Figure 7** Squared end-to-end distance as a function of the bead-number per chain. The symbols indicate the results from the standard bead-spring simulations obtained from the literatures[33,37,39,40]. The solid curves are the MCSS results with various $e^{v/kT}$ values. The error bars exhibit the standard deviation of the distribution. Black and yellow bars are for the bead-spring and the MCSS simulations, respectively.

Figure 8 left panel shows the linear relaxation modulus, $G(t)$, calculated by the Green-Kubo formula. As established, $G(t)$ shows the Rouse relaxation in a short time followed by a shoulder due to the retarded relaxation induced by the entanglement. In the right panel, to discuss the deviation from the Rouse relaxation, we plot $G(t)t^{1/2}$ following the earlier studies[16,35]. The MCSS results nicely overlap in a long-time region with those reported by Likhtman et al.[35] for $N_{KG}$ =100, 200, and 350, indicated by filled symbols. Meanwhile, our results are not compatible with those reported by Takahashi et al.[37] for $N_{KG}$ =100, 200 shown by unfilled circle and square. For $N_{KG}$ =512, the MCSS results locate between the curves by Cao and Likhtman[41] (cross) and Anwar and Graham[42] (triangle). In a short-time range, the MCSS results deviate upward from the bead-spring one, due to the cut-off induced by the coarse-graining. The deviation disappears within a certain duration, which is less than one order of magnitude in time.

The conversion factors for the modulus in Figure 8, $G_{KG}/G$, are listed in Table I. (Here, $G_{KG}$ is the unit of modulus for the bead-spring model.) We have found the empirical relation written as

$$G_{KG}/G = 0.014_3 N_{eSS}^\infty \qquad (13)$$

Eq (9) gives $G \propto \rho/N_{eSS}^\infty$ (here, the segment density $\rho$ = 4 is constant). Because this dependence is similar to that of the plateau modulus, the result implies that the plateau modulus is the characteristic modulus of the MCSS model, and the characteristic unit of the coarse-graining is $N_{eSS}^\infty$. We note that this result is not fully consistent with the results in Table I, where most of the conversion factors are related to $N_c$ rather than $N_{eSS}^\infty$.

We expect that the long-time behavior is dominated by the entanglement segment $N_{eSS}^\infty$. Based on this thought, the length, energy, and time units scale as $b' = b\sqrt{N_{eSS}^\infty}$, $kT$, and $\tau' = \tau N_{eSS}^{\infty\,2}$. The polymerization index and the segment density are written as $N' = N/N_{eSS}^\infty$ and $\rho' = \rho/N_{eSS}^\infty$. We may call these units as coarse-grained units. Because $N_{eSS}^\infty$ is in inverse proportional to the fugacity $\xi \equiv e^{\nu/kT}$ as seen in Fig 1 and eq 12, the coarse-grained units depend on the fugacity $\xi$. Let us consider two systems with different fugacity $\xi_1$ and $\xi_2$. These systems have different values of $N_{eSS}^\infty$, and the ratio between $N_{eSS,1}^\infty$ and $N_{eSS,2}^\infty$ is $\alpha = N_{eSS,1}^\infty/N_{eSS,2}^\infty = \xi_2/\xi_1$. Using $\alpha$, we can express the conversion factors between two systems as

$$\frac{b'_1}{b'_2} = \alpha^{1/2}, \qquad \frac{\tau'_1}{\tau'_2} = \alpha^2, \qquad \frac{N'_1}{N'_2} = \alpha^{-1} \qquad (14)$$

The results shown in Table I and Figs 1 and 6 demonstrate that these relations are reasonably fulfilled for small $e^{\nu/kT}$ values, where $N_{eSS}^\infty > 2$. However, when the slip-springs are densely introduced, the conversion factors behave differently. For the modulus, conversion between different systems is not straightforward, because the modulus depends on the magnitude of fluctuations imposed to the system[43–45]. If we assume that the magnitude of fluctuations is independent of $e^{\nu/kT}$, we have $G' = \rho'kT = G/N_{eSS}^\infty$. This relation is consistent with the results in Table I and eq 13.

From the above discussion, we may conclude that the dynamics of MCSS with different $e^{\nu/kT}$ is universal when $N_{eSS}^\infty > 2$, whereas the universality does not hold when the slip-springs are densely introduced. Uneyama[21,46] reported that the dynamics of single-bead systems with virtual springs are qualitatively modified when the density of virtual springs becomes large. MCSS exhibits a similar feature.

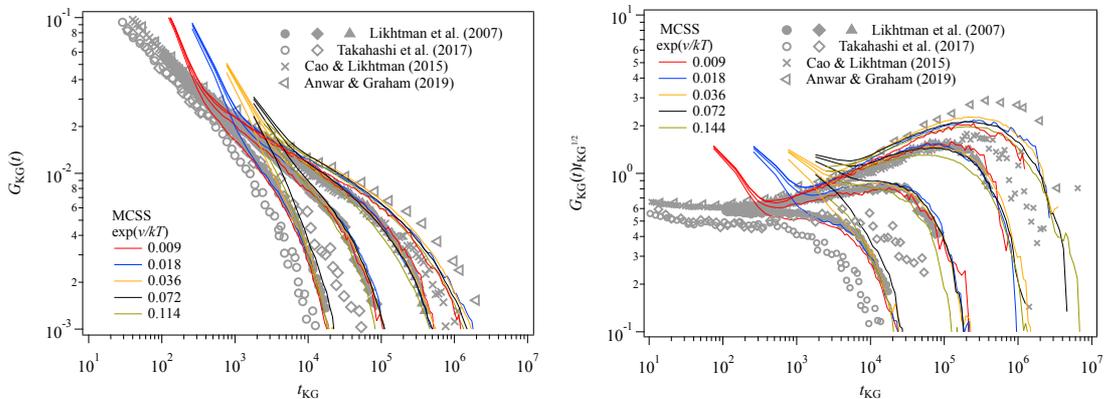

**Figure 8** Linear relaxation modulus $G(t)$ (left) and the reduced plot eliminating the Rouse relaxation $G(t)t^{1/2}$ (right) for $N_{KG} =$ 100, 200, 350, and 512, from left to right. Symbols show the literature

data for the bead spring simulations[35,41,42,47], and solid curves indicate the MCSS results with various $e^{v/kT}$ values. The conversion factors used are summarized in Table I.

Owing to the reasonable agreement with the bead-spring results for the dynamics at equilibrium, we compare the dynamics under shear. Figure 9 shows the flow curve for steady shear. Horizontal broken lines indicate the zero-shear viscosity calculated from $G(t)$. The MCSS results exhibit reasonable shear thinning, and the viscosity is consistent with the linear viscoelasticity in the low shear rate regime. However, under fast shear with $\dot{\gamma}_{KG} \gtrsim 10^{-4}$, the MCSS results deviate from the bead-spring data showing insufficient shear thinning. This deviation is due to the model construction of MCSS, for which the lowest viscosity must correspond to that for the Rouse chain without slip-springs. Because the Rouse chains do not exhibit shear thinning, the observed shear thinning is induced by the slip-springs. Thus, when the number of slip-springs becomes small under fast shear as shown below, the chain dynamics approaches to the Rouse behavior. Note that similar Rouse-like viscosity under fast shear flows has been also observed in the single-chain slip-spring model[48]. In the fast shear region, dashed lines indicate the Rouse viscosities for $e^{v/kT} = 0.009$, and the corresponding MCSS results (red curves) asymptotically approach those values.

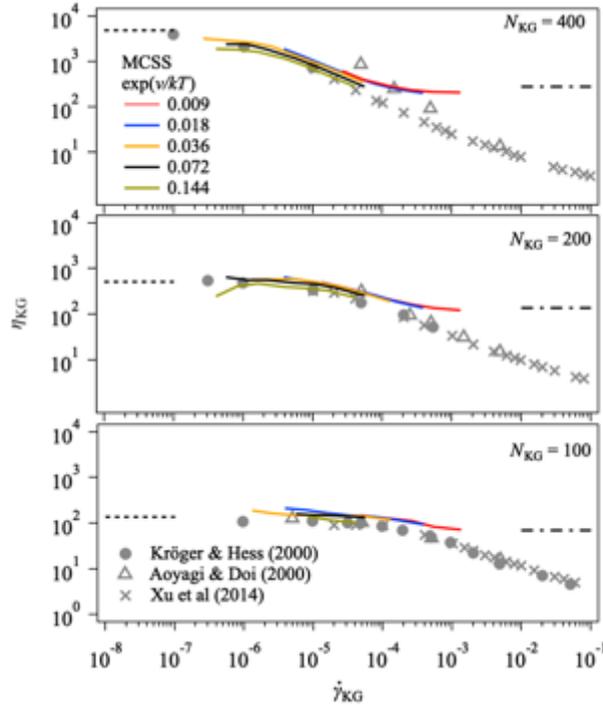

**Figure 9** Steady shear viscosity as a function of shear rate in comparison to the literature data for the standard bead-spring simulations with $N_{KG}$=400, 200 and 100, from top to bottom. Solid curves and symbols are for the MCSS results and the literature data for the bead-spring simulations[49–51], respectively. Broken lines in the low shear region indicate the zero-shear viscosity calculated from $G(t)$. Dashed lines in the high shear region corresponds to the Rouse viscosity for $e^{v/kT} = 0.009$. The conversion factors used are summarized in Table I.

For $N_{KG}$=400 and 200, Xu et al.[51] elaborated to extract the entanglement density by the Z1 code for the snapshots under steady shear. Figure 10 shows the comparison to their results for the shear induced reduction of the entanglement density. Interestingly, dynamic similarity is seen among the MCSS models with various $e^{\nu/kT}$ as well as the viscosity. The universal MCSS results are qualitatively similar with the bead-spring results, in a sense that the entanglement density decreases with the increasing shear rate in a power-law manner. However, the shear rate from which the reduction of entanglement occurs is significantly lower for MCSS than that for the bead spring simulations. For the MCSS simulations, the critical shear rate is close to the reciprocal longest relaxation time (i.e., the Weissenberg number $Wi_d \sim 1$). The reason for the discrepancy from the bead-spring simulations is unknown. A possible reason is that the constraint by slip-springs (or slip-links) is too strong compared with that by short-range Lennard-Jones repulsion in the bead-spring model. A slip-spring connects two segments and cannot be destroyed unless one of its ends is at chain ends. The reduction of the stress is directly correlated to the reduction of slip-springs, in the MCSS model. On the other hand, some entanglements detected by the primitive path extraction can be destroyed by local rearrangement of bead-spring chains. Then (at least a part of) the stress would be released by the local rearrangement, and thus the reduction of the entanglements under flow is somewhat suppressed in the bead-spring model. Meanwhile, we note that our result is consistent with the earlier atomistic molecular dynamics simulation[52] and the multi-chain slip-link[53] and slip-spring[26] simulations.

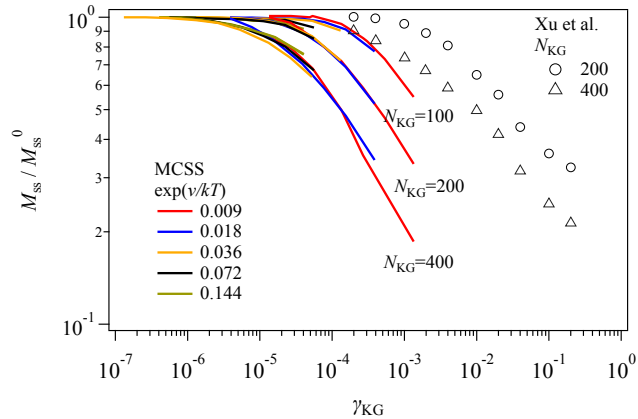

**Figure 10** Entanglement density normalized by the equilibrium value as a function of the shear rate for $N_{KG} = 200$ and 400. The symbols are the results from the bead-spring simulations extracted from the literature[51]. Solid curves indicate the MCSS results with various $e^{\nu/kT}$ values.

Figure 11 shows the comparison for the viscosity growth curve for $N_{KG} = 512$. MCSS overestimates the viscosity of the bead-spring model[41,42], not only for the steady-state seen in Figure 9 but also for the entire transient behavior. This discrepancy might be due to the construction of the MCSS model, for which the governing equations are derived for the dynamics under equilibrium. (The dynamics model of the MCSS is based on the detailed-balanced condition. The detailed-balanced condition ensures the relaxation to the thermal equilibrium in absence of flow, and physically reasonable in and near equilibrium. Its validity is generally not guaranteed under relatively fast flows.) The other coarse-grained models, such as the tube model[41,42] and the slip-link model[54], exhibit similar predictions, and the magnitude of discrepancy is smaller. This flaw of the MCSS model must be improved. In addition, the MCSS results for $e^{\nu/kT} \geq 0.072$ significantly smaller than those for the small $e^{\nu/kT}$ values. This result implies that the non-linear response of the MCSS model alters at $N_{eSS} \sim 2$. Namely, when $e^{\nu/kT} < 0.072$, i.e., $N_{eSS} > 2$, the universality is held among the MCSS models even under fast shear. Meanwhile, for $e^{\nu/kT} \geq 0.072$ and $N_{eSS} < 2$, the non-linear response under fast shear depends on $e^{\nu/kT}$ and not universal.

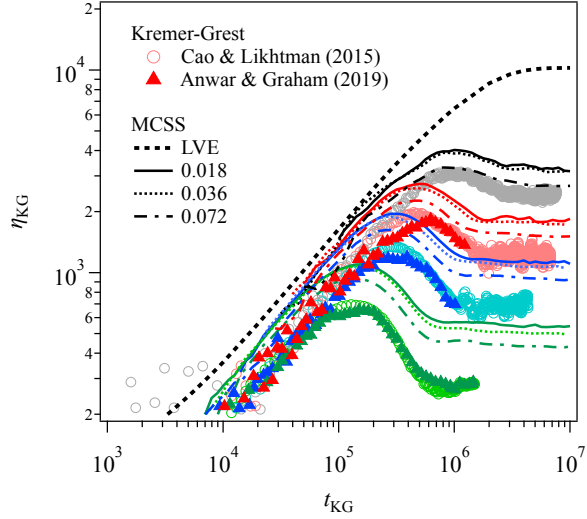

**Figure 11** Viscosity growth curve for $N_{KG}$=512 under fast shear with the Weissenberg number of 0.46, 1.15, 2.3 and 6.9, shown in black, red, blue and green, respectively. Circles and triangles are the bead-spring simulation results extracted from the literature. Tick dotted curve is the linear viscoelastic envelope determined from $G(t)$. Thin curves are the MCSS results with various $e^{\nu/kT}$ values.

For $N_{KG}$=512, Anwar and Graham[42] reported the transient behavior of the tube contour length under shear. Their definition is the sum of the end-to-end distance of subchains with a fixed bead number ($N$ = 65). Figure 12 indicates their data in comparison to the MCSS results for the Weissenberg number of 2.3 and 6.9. For the MCSS simulations, the tube contour length is straightforwardly defined as a path connecting the anchoring points of the slip-springs along the chain. Despite the difference in definition, the results agree with each other in some extent. In particular, the peak position is almost the same. For the MCSS simulations, this quantity roughly exhibits self-similarity, as well as the viscosity and the entanglement density.

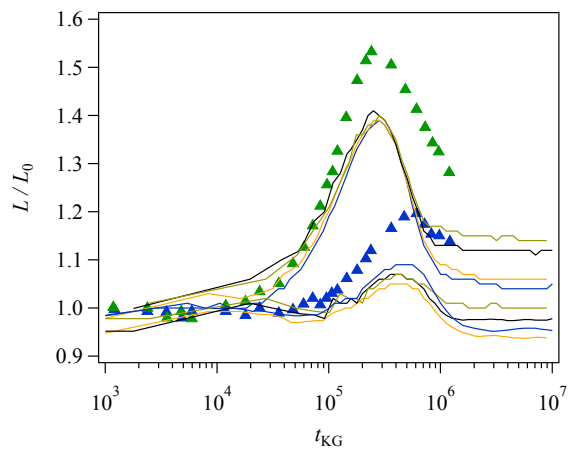

**Figure 12** Time development of the tube contour length $L$ normalized by the equilibrium value $L_0$ under start-up shear flows with the Weissenberg numbers of 6.9 (top) and 2.3 (bottom). The triangles show the results of the bead spring-simulations extracted from the literature. Solid curves indicate the MCSS results with various $e^{\nu/kT}$ values.

**CONCLUSIONS**

We performed a series of MCSS simulations for the polymer dynamics under equilibrium and shear flows. We found that the diffusion and the linear viscoelasticity are universal including the molecular weight dependence for the MCSS simulations with various slip-spring fugacity, $e^{\nu/kT}$. The results are consistent with those obtained for the standard bead-spring simulations if the $e^{\nu/kT}$-dependent conversion factors are adequately chosen for the units of molecular weight, length, time and modulus. These results demonstrate that the level of coarse-graining for the MCSS model can be arbitrary chosen, as expected from the model construction where the Rouse chains are connected via linear springs. However, the $e^{\nu/kT}$-dependence of the conversion factors cannot be described by the number density of slip-springs when the chains are densely connected with each other by the slip-springs. Under mild shear, the MCSS reasonably reproduce the bead-spring results, which include shear thinning. However, under fast shear, the MCSS overestimate the viscosity, due to the model construction, in which the viscosity is lower-limited by the value for the Rouse chain without slip-springs. In addition, the universality among the MCSS models does not hold for the high $e^{\nu/kT}$ values, where the chains are densely connected.

Even for the dynamics under equilibrium, the universality is not trivial for other systems like branch polymers, systems with molecular weight distribution, and polymer blends and copolymers. To deal with polymer solutions and other fancy systems, the DPD version[24,55,56] must be also examined. The studies toward such directions are being conducted and the results will be reported elsewhere.

ACKNOWLEDGEMENT

YM thanks Prof. Hiroshi Watanabe and Prof. Daniel Read for helpful duscissons. This study is supported in part by Ogasawara foundation, and JST-CREST number JPMJCR1992, Grant-in-Aid (KAKENHI) number JP19H01861 and JP20H04636, and JST-PRESTO number JPMJPR1992.

TOC graphics

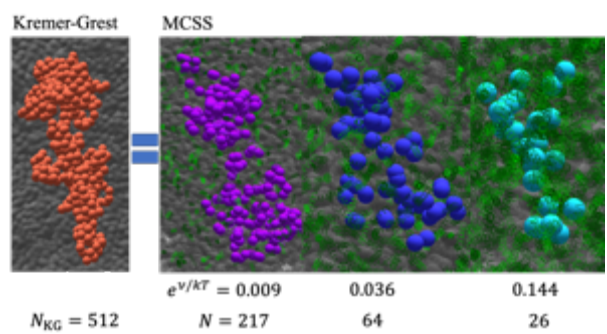